\documentclass[reprint,amsmath,amssymb,aps,prb,twocolumn,showpacs]{revtex4}

\makeatletter
\usepackage{graphicx}
\usepackage{dcolumn}
\usepackage{bm}
\usepackage{amsmath,graphicx,latexsym}
\usepackage{verbatim}   

\def\P{\mathbf{P}}
\def\D{\mathbf{D}}

\def\E{\bm{\mathcal E}}

\def\PTO{PbTiO$_3$}
\def\STO{SrTiO$_3$}

\def\Tx{\theta_x}
\def\Ty{\theta_y}
\def\Tz{\theta_z}
\def\To{\theta_0}

\def\bh{\boldsymbol{\theta}}

\def\Mc{$\mathcal{M}_{\rm C}$}
\def\Ma{$\mathcal{M}_{\rm A}$}
\def\Mb{$\mathcal{M}_{\rm B}$}

\def\bea{\begin{eqnarray}}
\def\nn{\nonumber\\}
\def\eea{\end{eqnarray}}
\def\beq{\begin{equation}}
\def\eeq{\end{equation}}

\begin{document}

\title{Electrically driven octahedral rotations in SrTiO$_3$ and
PbTiO$_3$}

\author{Jiawang Hong}
\email{hongjw10@physics.rutgers.edu}
\affiliation{ Department of Physics and Astronomy, Rutgers University,
 Piscataway, NJ 08854-8019, USA }

\author{David Vanderbilt}
\affiliation{ Department of Physics and Astronomy, Rutgers University,
 Piscataway, NJ 08854-8019, USA }

\date{\today}

\begin{abstract}

We investigate the oxygen octahedral rotations that occur
in two perovskites, \STO\ and \PTO, as a function of applied
three-dimensional electric displacement field, allowing us to
map out the phase diagram of rotations in both the paraelectric
and ferroelectric regions of the polar response.
First-principles calculations at fixed electric displacement
field are used to extract parameters of a Landau-Devonshire
model that is analyzed to identify the phase boundaries
between different rotational states.  The calculations reveal
a rich phase diagram of rotations versus applied field in
both \STO\ and \PTO, although the details are quite different
in the two cases.

\end{abstract}


\maketitle

\marginparwidth 2.7in
\marginparsep 0.5in
\def\dvm#1{\marginpar{\small DV: #1}}
\def\jhm#1{\marginpar{\small JH: #1}}
\def\scr{\scriptsize}

\section{Introduction}
\label{sec:intro}

Oxygen octahedral rotations can have a
significant impact on the behavior of ABO$_3$ perovskites,
affecting electronic, dielectric, ferroelectric and magnetic properties.
For example, the octahedral rotations couple strongly with
the magnetic structure in transition-metal perovskites by
modifying
the metal-oxygen-metal bond angles that are critical to
determine the magnetic
interactions.\cite{Nicole,spin} In some materials, such as
SrTiO$_3$, the oxygen rotations give rise to a non-polar
antiferrodistortive (AFD) ground state contribute to the
suppression of ferroelectric (FE) order.\cite{Zhong,Sai,Eklund}.
However, the recent discovery of rotation-driven improper ferroelectricity
in a superlattice~\cite{Eric} has inspired a search for
this type of ferroelectricity in other types of
materials.~\cite{Nicole,Ale,Jorge,Claude} Because they can also couple
with magnetic properties, octahedra rotations offer a
promising approach to the discovery or design of new
multiferroic perovskites.~\cite{Nicole, Lawes,Fennie,James1,James2}  

In the cubic structure, \STO\ and \PTO\ both show
an AFD instability at the zone corner ($R$ point) of the
Brillouin zone.\cite{Ghosez1,Ghosez2}  Following this instability
leads to the experimentally observed tetragonal ground state
with rotations along the [001] axis for \STO, while \PTO\ prefers
rotations along the [111] axis.
For \PTO, however, this is not the equilibrium structure; instead,
a strong FE instability at $\Gamma$ out-competes the AFD instability,
giving rise to a tetragonal FE ground state without rotations.
Nevertheless, the AFD modes provide a potential source
of instability in \PTO, as has been predicted for example
for surface~\cite{Munkholm,Karin} and interface~\cite{Eric} 
geometries.  (In \STO, a weak FE instability at $\Gamma$ is
found in some calculations, depending sensitively on lattice
constant and other details.  However, experimentally the material
just barely avoids this instability, remaining paraelectric
down to zero temperature.)

While oxygen octahedral rotations clearly play an important role
in these and other perovskites, and while they are known to be
strongly affected by stress~\cite{James1,James3,May} and
temperature,\cite{Muller} we are not aware of any previous
study of the effects of \textit{electric} fields on the
AFD rotations.  In this paper, we study the phase
transition behavior of the AFD modes in \STO\ and \PTO\  under 
three-dimensional constant electric displacement field.

Of course, \STO\ and \PTO\ are qualitatively different in that
the latter is ferroelectric while the former remains paraelectric
down to zero temperature.
However, the choice of fixed electric displacement field $\D$
for the boundary conditions in this study allows us to treat both
materials on an equal footing.  The situation would have been
much more complicated if we had chosen to work at fixed electric
field $\E$, because the energy
landscape is multivalued and the paraelectric configuration is
unstable at small $\E$ in a ferroelectric material like \PTO.
At fixed $\D$, however, the energy landscape remains single-valued,
thus allowing access to the entire electric equation of state
for \PTO\ as well as \STO.\cite{Max-NM,fixd}  Indeed, both materials
have a large static dielectric constant, so that mapping at
fixed $\D$ is qualitatively similar to mapping at fixed $\P$.
Thus, in our study, the main qualitative difference between
\STO\ and \PTO\ will be related to the fact that the rotational
instability that prefers to develop along a [001] axis in
\STO\ instead prefers a [111] axis in \PTO.

The paper is organized as follows.  In Sec.~\ref{sec:prelim}
we introduce the Landau-Devonshire model, provide the details
of our first-principles calculations, and specify the terminology
for symmetries that will be used later.
In Sec.~\ref{sec:result} we present the results of the first-principles
calculations in one-dimensional $\D$-field space and discuss the
fitting of the model, which is then used to compute the phase
diagram of rotational phases in three-dimensional $\D$ space.
Finally, Sec.~\ref{sec:summ} contains a summary.

\section{Preliminaries}
\label{sec:prelim}

\subsection{Landau-Devonshire model}
\label{sec:model}

In order to explore the octahedral-rotation phase diagram in the
space of $\D$ fields, the internal energy $U$ has to be
calculated and minimized on a three-dimensional mesh of
$\D$ values.  Near the phase boundaries between different
rotational phases, this process would be quite tedious;
the first-principles calculations would need to be very
carefully converged, and the procedure would become
quite time-consuming.  We therefore introduce a
Landau-Devonshire model to study the phase transitions in this
system, with the coefficients in the model being obtained from
fitting to our first-principles results on a smaller database
of $\D$ values.  This model can then be used to locate the
phase boundaries efficiently.

As mentioned above, the dominant AFD rotational instabilities
for paraelectric \STO\ and \PTO\ are both at the $R$ point
in the Brillouin zone (corresponding, in the most general
case, to the $a^-b^-c^-$ Glazer notation).  We therefore focus
on these modes here, and define a vector
octahedral rotation $\bh=(\Tx,\Ty,\Tz)$ describing a rotation
by angle $\Tx$ around the $x$-axis, etc.~(or, more generally,
by angle $\theta=|\bh|$ about axis $\hat{\theta}$).

Within our Landau-Devonshire model, then, the internal energy
$U_{\rm tot}(\D,\bh)$ is expanded
as a function of displacement field $\D=(D_x,D_y,D_z)$ and
octahedral rotations $\bh=(\Tx,\Ty,\Tz)$ as
\begin{equation}
U_{\rm tot}= U_D+U_\theta+U_{\rm int}
\label{eq:Utot}
\end{equation}
where
\begin{widetext}
\begin{eqnarray}
U_D(\D) &=& \alpha (D_x^2+D_y^2+D_z^2) + \beta
  (D_x^4+D_y^4+D_z^4) + \gamma (D_x^2 D_y^2 +D_x^2 D_z^2 + D_y^2
  D_z^2) ,
\label{eq:UD}\\
U_\theta(\bh) &=&
   \mu (\Tx^2+\Ty^2+\Tz^2)+ \omega (\Tx^4+\Ty^4+\Tz^4)+
    \sigma (\Tx^2 \Ty^2 + \Tx^2 \Tz^2 + \Ty^2 \Tz^2) ,
\phantom{\int} 
\label{eq:Utheta}\\
U_{\rm int}(\D,\bh) &=&
   \tau (\Tx^2 D_x^2 + \Ty^2 D_y^2 + \Tz^2 D_z^2) +
    \lambda ( \Tx^2 D_y^2 + \Tx^2 D_z^2 +  \Ty^2 D_x^2 + \Ty^2 D_z^2
    + \Tz^2 D_x^2 + \Tz^2 D_y^2 )  \nn
    &&\quad + \kappa (\Tx \Ty D_x D_y + \Tx
    \Tz D_x D_z + \Ty \Tz D_y D_z) .
\label{eq:Uint}
\end{eqnarray}
\end{widetext}
Here we have made the approximation of truncating the expansion
systematically at overall fourth order, and $\alpha, \beta,
\gamma, \mu, \omega, \sigma, \tau ,\lambda$ and $\kappa$ are
coefficients that need to be fitted from the first-principles calculations.
The terms in $U_{\rm int}$ describe the coupling of $\D$ and octahedral
rotations. There is no strain in this expansion since each term
is defined assuming that the strain is fully relaxed for each
$(\D,\bh)$ value.

In the present work we are really only interested in the
internal energy difference
$U_{\rm tot}(\D,\bh)- U_{\rm tot}(\D,0)$
between the states with and without octahedral
rotations.  We denote this quantity simply as
$U$ and note that
\begin{equation}
U(\D,\bh)=U_\theta(\bh)+U_{\rm int}(\D,\bh) .
\label{eq:U}
\end{equation}

In order to fit the coefficients from first-principles
calculations, we first apply $\D$ along just one Cartesian
direction, which we choose as $D_z$, to find the coefficients  
$\mu, \omega, \sigma, \tau$ and $\lambda$.
For this case we set $D_x=D_y=0$ and find
\begin{eqnarray}
U \!&=&\!  \mu \theta^2
   + \tau  D_z^2  \Tz^2
   +\lambda  D_z^2 ( \Tx^2 + \Ty^2 )
\nonumber\\&&
   +\omega \theta^4
   +(\sigma-2 \omega) (\Tx^2 \Ty^2 + \Tx^2 \Tz^2 + \Ty^2 \Tz^2) .
    \label{eq:modelz}
\end{eqnarray}
We then do a series of calculations in which 
we choose different initial structures with equillibrium rotations along
$\hat{\bh}$=[100], [110], [001] or [111] at $D_z$=0.0\,a.u.,
and for each choice (and for each $D_z$) we relax all the
coordinates to obtain the internal energy $U(D_z,\hat{\bh}$).
(We increase $D_z$ in increments of 0.04\,a.u.\ up to 0.12\,a.u.\ for
\STO, and increments of 0.02\,a.u.\ up to 0.08\,a.u.\ for \PTO.)
Fitting the model parameters to this first-principles
database of information, we obtain all the coefficients in
Eqs.~(\ref{eq:Utheta}-\ref{eq:Uint}) except for $\kappa$.
We then do one more series of calculations with both
$\D$ and  $\bh$ along the $[111]$ direction,
i.e., $\D=(D_0,D_0,D_0)$ and $\bh=(\To,\To,\To)$, for which the
model predicts
\beq
  U= 3 \mu \To^2+3(\omega+\sigma) \To^4 +3(\tau + \kappa +
  2\lambda) \D_0^2 \To^2 .
  \label{eq:model111}
\eeq
(Here $D_z$ is increased in steps of 0.02\,a.u.\ up to 0.08\,a.u.\ for
both materials.)
Fitting in a similar way to these results, we obtain
the parameter $\kappa$ as well.
Once all the parameters are in hand, we can
go back to Eq.~(\ref{eq:U}) and study
the full behavior of octahedral rotations as a function of
three-dimensional $\D$ space using this model. 

\subsection{First-principles methodology}
\label{sec:methods}

Our calculations were performed within density-functional theory
in the local-density approximation\cite{Perdew-Wang}
using norm-conserving pseudopotentials \cite{Norm-conserve} and a
plane-wave cutoff of 60 Ha. A 6$\times$6$\times$6
Monkhorst-Pack grid\cite{MP} was used to sample the Brillouin zone.
The unit cell for simulating the $R$-point rotation is doubled
to obtain a 10-atom fcc cell. The
atomic coordinates and lattice vectors of this cell
were relaxed until all atomic force components
were smaller than 10$^{-5}$\,Ha/Bohr and
all stress components were below 10$^{-7}$\,Ha/Bohr$^3$. We
used the open-source ABINIT code package\cite{Gonze} with
the implementation of the constant-displacement-field method in
3-dimensions~\cite{fixd} to calculate the internal energy
at a each specified $\D$ field.

\begin{figure*}
  \begin{center}
    \includegraphics{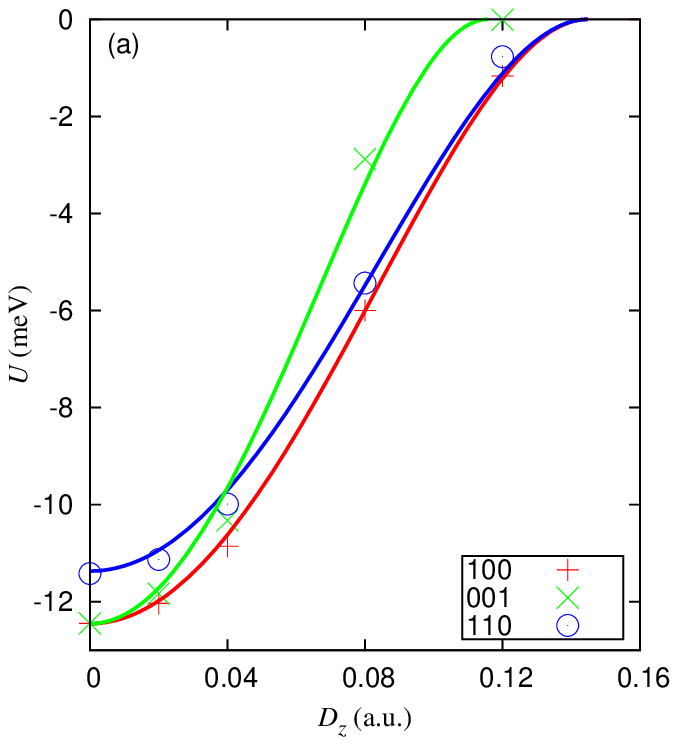}
    \includegraphics{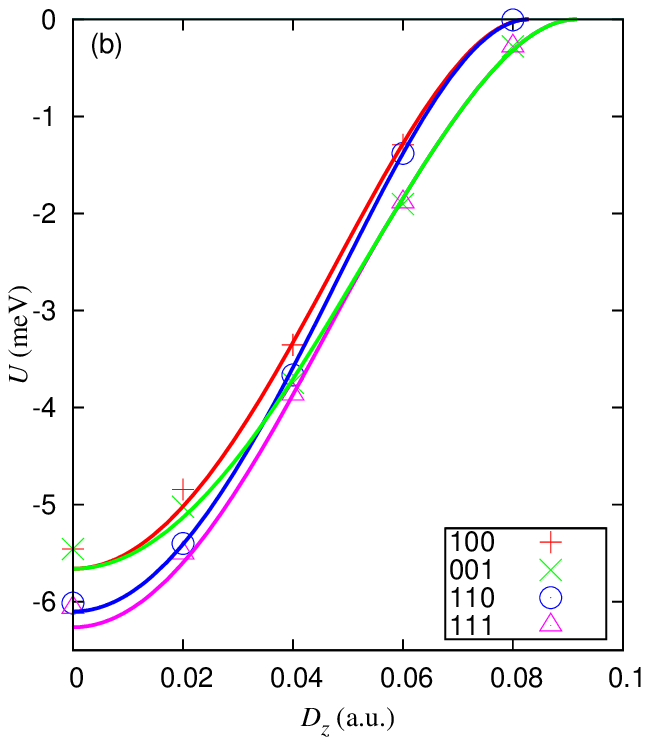}
  \end{center}
  \caption{Internal energy $U$ of Eq.~(\ref{eq:U}) for $\D$
  applied along the $\hat{z}$ direction, for phases with the
  octahedral rotations constrained to be about different axes
  as indicated in the legend.  (a) \STO; (b) \PTO. Symbols
  are from first-principles calculations; curves are from
  the Landau-Devonshire model.}
  \label{fig:energy}
\end{figure*}

\subsection{Terminology for symmetries}
\label{sec:symm}

Here we introduce the notations that we will use for describing
rotational phases, following a similar scheme as the one often
uses for polarization.~\cite{david} When the octahedral rotation
axis is constrained to a symmetry axis lying along
$\langle001\rangle$, $\langle111\rangle$, or $\langle011\rangle$,
the resulting phase becomes
tetragonal ($\mathcal{T}$), rhombohedral ($\mathcal{R}$), or
orthorhombic ($\mathcal{O}$),
respectively.~\cite{note1}
Similarly, the $\mathcal{M}$ phases arise when
rotation axis is confined to a mirror plane. There are three
cases: \Mc, in which the axis is along $[0,u,v]$; and \Ma\ or \Mb, in
which the axis is along $[uuv]$ with $u<v$ or $u>v$, respectively.
The triclinic phase (Tri) occurs if the axis is along $[uvw]$
with $u \neq v \neq w \neq 0 $. We also introduce the Cartesian
subscript $\alpha = \{x,y,z\}$ to specify the unique Cartesian
direction when needed.
For example, $\mathcal{T}_{\alpha}$ denotes the
tetragonal phase with rotation axis along direction $\alpha$,
while $\mathcal{O}_{\alpha}$ and $\mathcal{M}_{{\rm C}\alpha}$
denote the orthorhombic phase and \Mc\
phases with rotation axis lying in the plane perpendicular to
the $\alpha$ direction.  Similarly,
$\mathcal{M}_{{\rm A}\alpha}$ and $\mathcal{M}_{{\rm B}\alpha}$
are the \Ma\ and \Mb\ phases with the non-equal component $v$
in $[uuv]$ along the $\alpha$ direction.

\section{Results}
\label{sec:result}

\subsection{First-principles calculations}

\begin{figure}
  \begin{center}
    \includegraphics{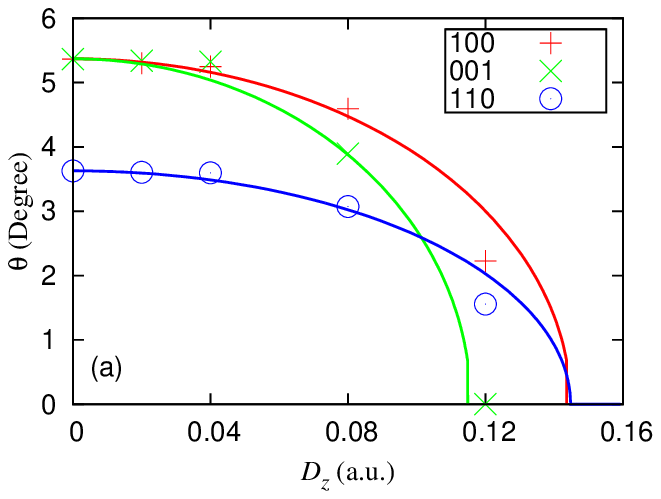}
    \includegraphics{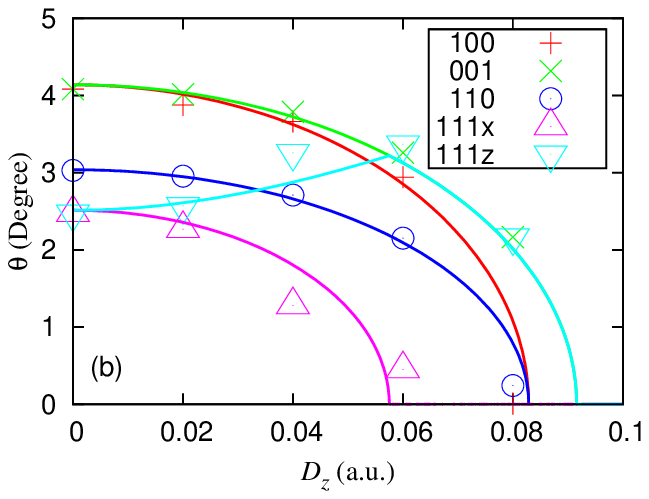}
  \end{center}
  \caption{Octahedral rotation angles $(\theta_x,\theta_y,\theta_z)$
  for different phases for $\D$ applied along $\hat{z}$. (a)
  \STO; (b) \PTO.  For [110] cases ($\theta_x$=$\theta_y$),
  $\theta_x$ is plotted.  For the \PTO\ [111]-derived case
  ($\theta_x$=$\theta_y$$\ne$$\theta_z$), $\theta_x$ and $\theta_z$
  are plotted.  Symbols are from first-principles calculations;
  curves are from model.}
  \label{fig:angle}
\end{figure}

We first carry out a series of calculations, starting from $\D=0$
and increasing $D_z$ in steps of 0.04\,a.u.\ for \STO\ and
0.02\,a.u.\ for \PTO, to explore
the resulting behavior for the case that the
octahedral rotation is constrained to lie along the
[100], [110], or [001] axis.   At each $D_z$,
the structure is fully relaxed with respect to both ionic positions
and lattice parameters.  In all three cases in both materials,
the rotations, which
are fully developed at $D_z=0$, are gradually suppressed with
increasing $D_z$ until they disappear completely at a critical
value of $D_z$.  We also attempt this procedure for the case
that the octahedral rotations started along the [111] direction
at $\D=0$.  However, for \STO\ a rotation along [111] is
a saddle point, rather than a local minimum, of the $\D=0$ energy
landscape, and the breaking of the three-fold symmetry about [111]
by the applied $D_z$ immediately causes the rotation axis to
switch to either the [110] or [001] direction.
For \PTO, by contrast, the $\D=0$ system has its minimum-energy
AFD axis along [111], and we can also follow the evolution of
this fourth case as $D_z$ is applied.  In this case we
find that
$\theta_z$ gradually increases, and $\theta_x=\theta_y$ gradually
decrease, with increasing $D_z$, until a critical value is reached at
which $\theta_x$ and $\theta_y$ vanish and the solution merges
with the one with the rotation axis constrained to $[001]$.
The results of these calculations are shown as the symbols
in Figs.~\ref{fig:energy} and \ref{fig:angle}, where
the internal energy difference $U$ of Eq.~(\ref{eq:U}) and
the equilibrium rotation angles are plotted versus $D_z$.

We also carry out calculations for both materials with $\D$ and
the rotation axis both constrained to lie along [111].  As
mentioned in Sec.~\ref{sec:model}, the purpose of this is just
to obtain the additional coefficient $\kappa$ that was not
determined from the calculations with $\D$ along [001], so
it was not necessary to study other rotation axes for this
case.  The results (not shown) again indicate that the rotations
gradually decrease with increase of $D_0$ for $\D=(D_0,D_0,D_0)$,
although in the case of \STO\ the rotations never vanish over
the range of $D_0$ up to 0.10 a.u. studied here.

\subsection{Fitting of the model parameters}

%
We now use the results of the above first-principles calculations
to determine the parameters in Eqs.~(\ref{eq:Utheta}-\ref{eq:Uint})
following the procedure detailed at the end of Sec.~\ref{sec:model}.
The resulting parameter values are reported in Table~\ref{tab:coeff}.
The predictions of the fit (solid curves) are compared with the
direct first-principles results (symbols) in
Figs.~\ref{fig:energy} and \ref{fig:angle}.
It is clear that the model agrees quite well with the
first-principles calculations. 

\begin{table}[b]
  \centering
  \caption{Fitted coefficients of the Landau-Devonshire model of
  Eqs.~(\ref{eq:Utheta}-\ref{eq:U}), defined with energies in meV,
  rotation angles in degrees, and displacement fields in a.u.}
  \begin{ruledtabular}
  \begin{tabular}{lcccccc}
      & $\mu$ & $\omega$ & $\sigma$ & $\tau$ & $\lambda$ & $\kappa$ \\
      \hline
\STO  &  --0.863 & 0.015 & 0.036 & 64.45 & 41.19 & --139.36  \\
\PTO  &  --0.661 & 0.019 & 0.033 & 78.80 & 96.27 & --147.91  \\
  \end{tabular}
  \end{ruledtabular}
  \label{tab:coeff}
\end{table}

\subsection{Details for $\D$ along [001]}

 From Fig.~\ref{fig:energy}, we can see that \STO\ and \PTO\ have
different octahedral rotation patterns. At $\D=0$, \STO\ has
the lowest energy in the $\mathcal{T}$ phase, which is its true
ground state experimentally below 105\,K, and the highest energy
in the $\mathcal{R}$ phase (not shown in the figure because it
is destabilized by any finite $D_z$.)  In \PTO, on the other hand,
the energy ordering is just the opposite, with the $\mathcal{R}$
phase lowest and the $\mathcal{T}$ phase highest in energy. In
the context of Eq.~\ref{eq:modelz}, the energy ordering of the
phases at $\D=0$ is determined by the combination of parameters
($\sigma - 2 \omega$), with the  $\mathcal{R}$ or $\mathcal{T}$
phase lowest in energy when this combination is negative
or positive, respectively.  This is confirmed by the coefficients
in Table~\ref{tab:coeff}.

\begin{figure*}[<+htpb+>]
  \begin{center}
    \includegraphics[height=1.93 in]{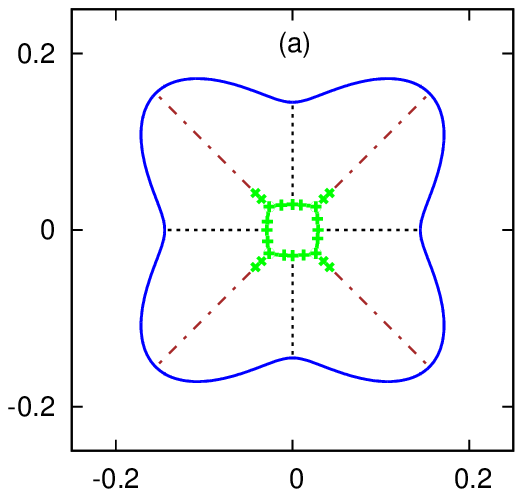}
    \includegraphics[height=1.93 in]{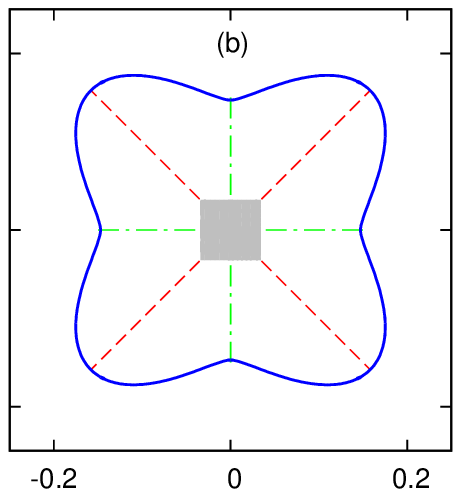}
    \includegraphics[height=1.93 in]{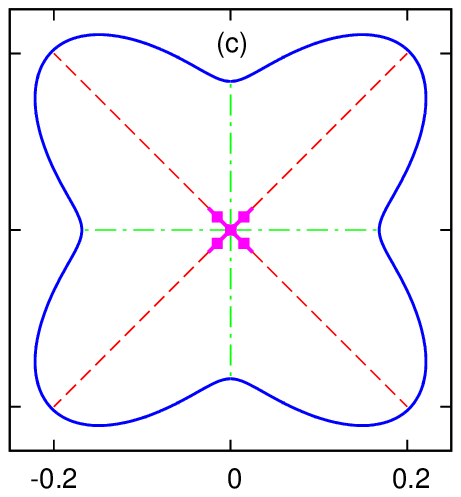}
    \includegraphics[height=1.93 in]{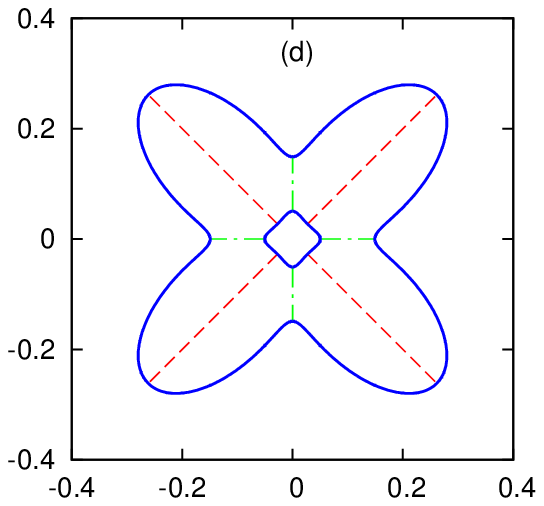} 
    \includegraphics[height=1.93 in]{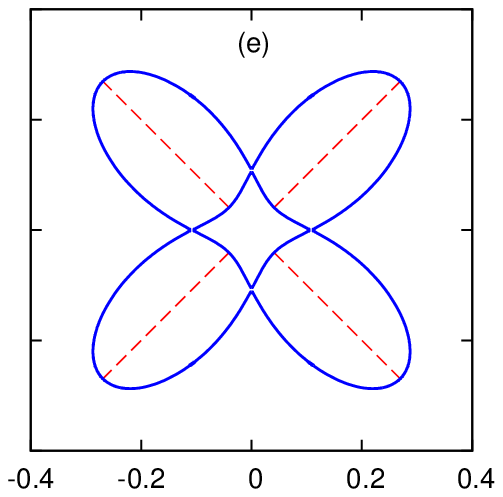} 
    \includegraphics[height=1.93 in]{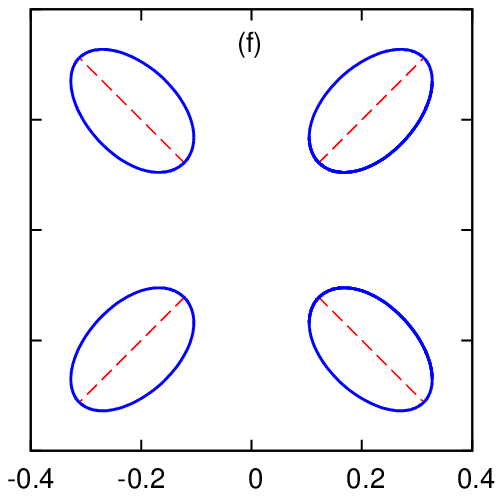} 
  \end{center}
  \caption{Phase diagram for rotational phases of \STO\ under
  applied $\D$ field.  Each panel is a cut plotted in the
  $D_x$-$D_y$ plane (note the change of scale between top
  and bottom panels) at fixed $D_z$.
  (a) $D_z$=0.00 a.u.; (b) 0.015 a.u.; (c) 0.08 a.u.; (d) 0.16
  a.u.; (e) 0.172 a.u.; (f) 0.24 a.u.  Solid lines are second-order
  boundaries; decorated solid lines are first-order boundaries;
  and dashed, dash-dotted, and dotted lines are special cases of
  higher symmetry induced by high-symmetry $\D$-vectors (see
  text for details). Gray area in (b) is detailed in
  Fig.~\ref{fig:blowup}.}
  \label{fig:sto}
\end{figure*}

As $D_z$ increases, Fig.~\ref{fig:energy} shows that the
internal energy $U$ of \STO\ and \PTO\ increases and finally
reaches zero. Recalling that $U$ is measured relative to the
structure with no rotations, we conclude that the octahedral
rotations disappear at a sufficiently
high $D_z$ field. However, the behavior is different
for these two materials.  Fig.~\ref{fig:energy}(a) shows
that for \STO\ the $\mathcal{T}$ phase\cite{note1}
with rotation axis along [100] or [010]
has the lowest energy as $D_z$ increases, while the
$\mathcal{T}$ phase with rotation along [001] increases sharply
in energy and becomes the least favorable state when $D_z > 0.05$\,a.u.
This suggests that the $\mathcal{T}$ phase with its rotation axis
perpendicular to $\D$ is favored, while the one with rotation
axis parallel to $\D$ is suppressed.
However, \PTO\ has a quite different behavior, as can be
seen in Fig.~\ref{fig:energy}(b).
The state of lowest internal energy at $\D=0$ is the
$\mathcal{R}$ phase. As $D_z$ increases,  the rotational
axis is perturbed to be along $[uuv]$ for
$v>u$, putting the system in the $\mathcal{M}_{\rm A}$ phase.
Eventually, the internal energy of this state
merges into the curve for the $\mathcal{T}$
phase (axis along [001]), indicating a phase
transition from $\mathcal{M}_{\rm A}$ to $\mathcal{T}$ at some
critical value of $D_z$.  The $\mathcal{T}$ phase, with its rotation
axis parallel to $D$, is then favored at higher $D_z$, until
there is a second phase transition at which the rotations disappear.

The details of the rotational behavior in \PTO\ can be seen more
clearly in  Fig.~\ref{fig:angle}(b), which shows the variation of
the various rotation angles with $D_z$ field. The rotation angles
decrease as $D_z$ increases for all phases except for the initial
$\mathcal{R}$ phase. This phase is immediately perturbed to become
\Ma\  as soon as a non-zero $D_z$ is present.  With increasing
$D_z$, the rotation angles $\theta_x$ and $\theta_y$ decrease,
but $\theta_z$ increases.  That is, the rotation axis starts from
[111] ($\mathcal{R}$) and then rotates in the (1$\bar{1}$0) plane
(\Ma) towards [001] ($\mathcal{T}$).  We can now see that the
critical $D_z$ at which $\mathcal{T}$ phase is reached (i.e., at
which $\theta_x$ and $\theta_y$ vanish) is at $D_z$=0.058\,a.u.
This also corresponds to the merger of \Ma\ and $\mathcal{T}$
phases in the internal energy curves of Fig.~\ref{fig:energy}(b).
For larger $D_z$, $\theta_z$ then decreases monotonically
and reaches zero at $D_z$=0.092\,a.u.

For \STO, on the other hand, the picture is simpler.  As $D_z$
increases in Fig.~\ref{fig:angle}(a), the rotation axis remains
along [100] while the amplitude of $\theta_x$ monotonically
decreases and disappears entirely at a critical value
of $D_z$=0.144\,a.u.

\subsection{Three-dimensional $\D$ field}

We now turn to a detailed discussion of the behavior of \STO\ and
\PTO\ as a function of three-dimensional $\D$ field, based on
the model of Eqs.~(\ref{eq:Utheta}-\ref{eq:U}) using the coefficients
fitted from first principles as reported in Table \ref{tab:coeff}.

First, note that because the coefficient $\mu$ is negative
in both \STO\ and \PTO, we are guaranteed to get a phase with
non-zero rotations at small $D$.  Also, because of the non-zero
value of $\kappa$, we generically obtain a triclinic rotational
axis ($\theta_x\ne\theta_y\ne\theta_z$) at a general point
$D_x\ne D_y\ne D_z\ne 0$ in $\D$ space.  High-symmetry phases
will only exist under special conditions, i.e., when one or
more $\D$ components vanish, or when two or more $\D$
components are equal.
 
\subsubsection{Phase diagram for \STO} \label{sec:sto-results}

Figure \ref{fig:sto} shows several two-dimensional $(D_x,D_y)$ slices of
the three-dimensional phase diagram of STO taken at different values
of $D_z$.
In these panels, the outer solid (blue) boundaries (and also the
inner ones in Fig.~\ref{fig:sto}(d-e)) indicate a second-order phase
transition from a phase with octahedral rotations to a phase without
rotations.  Other solid lines represent first-order phase boundaries
as will be explained below.
Dashed and dotted lines are not true phase boundaries, but
instead denote high-symmetry structures that occur as special
cases along special lines or planes in $\D$ space; we use
dashed lines (red) for \Ma\ or \Mb\ phases, dotted lines
(black) for $\mathcal{T}$ phases,  dashed-dotted lines
(green) for \Mc\ phases, and short-dashed-dotted lines
(brown) for $\mathcal{O}$ phases, using the notation developed
in Sec.~\ref{sec:symm}.

\begin{figure}
  \begin{center}
    \includegraphics[width=1.6 in]{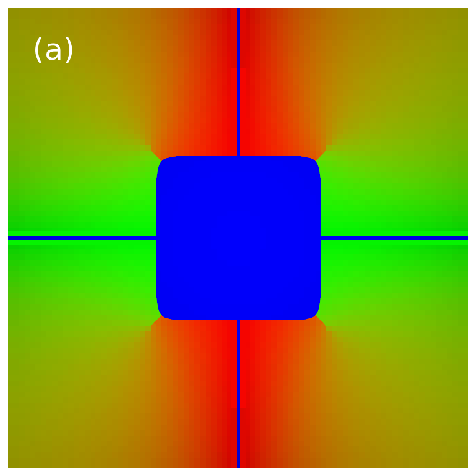}
    \includegraphics[width=1.6 in]{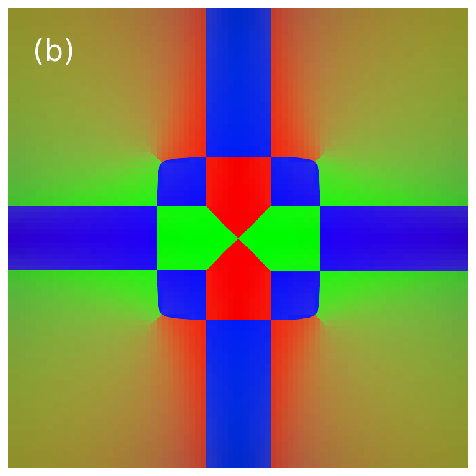}
    \includegraphics[width=1.6 in]{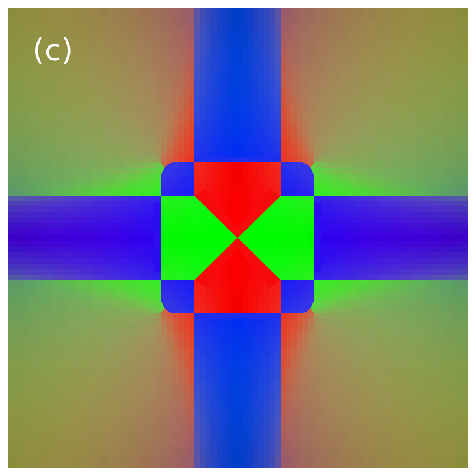}
    \includegraphics[width=1.6 in]{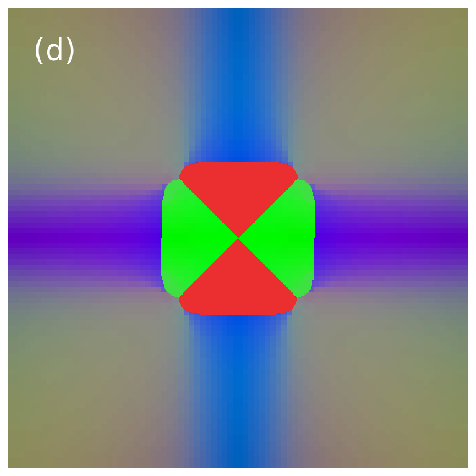}
    \includegraphics[width=1.6 in]{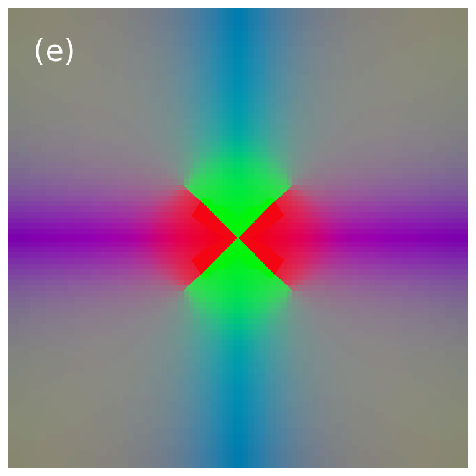}
    \includegraphics[width=1.6 in]{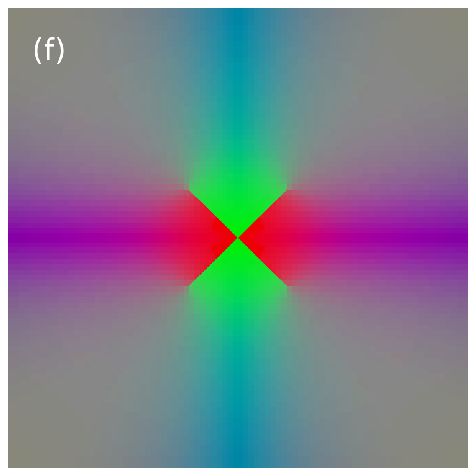}
  \end{center}
  \caption{Enlargement of the central portion of Fig.~\ref{fig:sto}
      ($D_x, D_y\in[-0.08,0.08]$\,a.u.) for small $D_z$ values.
      (a) $D_z$=0.00 a.u.; (b) 0.01 a.u.; (c) 0.015 a.u.; (d) 0.025
      a.u.; (e) 0.04 a.u.; (f) 0.05 a.u.  Color coding is such
      that pure red, green and blue correspond to the $\mathcal{T}_x$,
      $\mathcal{T}_y$ and $\mathcal{T}_z$ phases respectively,
      with color values weighted according to $|\theta_x|$,
      $|\theta_y|$ and $|\theta_z|$ for intermediate phases.}
  \label{fig:blowup}
\end{figure}

In Fig.~\ref{fig:sto}(a), for $D_z=0$, the
squarish solid curve marked by crosses (green) is a first-order
boundary separating the
$\mathcal{T}_z$ phase (inside) from phases with
$\theta_z$=0 (outside).  For generic ($D_x,D_y)$ outside, this
corresponds to the \Mc$_z$ phase (recall, from Sec.~\ref{sec:symm},
that this means that the rotation axis lies in the $\theta_x-\theta_y$
plane).  Along the horizontal axis ($D_y=0$) outside, shown
by the dotted (black) line, the
$\mathcal{T}_z$ and $\mathcal{T}_y$ phases are degenerate.  However,
any small finite $D_y$ favors the $\mathcal{T}_y$ phase and
adds a small $\theta_x$ component (via the $\kappa$ term) so that
the \Mc$_z$ phase results.  That is, crossing this dotted line
from negative to positive $D_y$ just causes $\theta_x$ to cross
smoothly through zero, so that this is not a true phase boundary.

Along the [110] direction in $\D$ space, the behavior is rather
complex.  We let $D_x=D_y=D_0$.  Recall that inside the square
region (small $D_0$) one finds the $\mathcal{T}_z$ phase.
Next comes a segment of first-order phase boundary, indicated again
by a solid line with crosses (green), along which there are two
degenerate \Mc$_z$ phases with rotation angles
($\theta_a$,$\theta_b$,0) and ($\theta_b$,$\theta_a$,0)
(with $\theta_a \neq \theta_b$).  Any small step away from this
line (while remaining in the $D_z$=0 plane) favors one or the
other of these phases, and also slightly perturbs its angles
$\theta_x$ and $\theta_y$.  Thus, when crossing this line, both
$\theta_x$ and $\theta_y$ jump discontinuously.
When $D_0$ increases further, as shown by the dash-dotted (brown) line,
one finds the $\mathcal{O}_z$ phase exactly on this line, but it
is just a special case of the \Mc$_z$  phase as $|\theta_x|$
and $|\theta_y|$ cross smoothly through each other.
Like the dotted (black) line, therefore, this is not a true phase
boundary.

As $D_z$ increases from zero, the behavior of the phase diagram
is initially very complex, especially in the vicinity of the
squarish central region of Fig.~\ref{fig:sto}(b).  The phase
behavior in the outer region at $D_z=0.015$\,a.u.\ is shown in
Fig.~\ref{fig:sto}(b).
The \Mc$_z$ phase at $D_z$=0 is perturbed to become triclinic as
$\theta_z$ becomes
non-zero linearly in $D_z$.
The $\mathcal{T}_x$ and $\mathcal{T}_y$ lines at $D_z$=0 are perturbed
to \Mc\ structures as shown by the dash-dotted
(green) lines, and the $\mathcal{O}_z$ lines are converted to
\Ma\ and \Mb\ structures as shown by the dashed (red) lines.
There are no true phase transitions when crossing these non-solid
lines.

The phase behavior in the
inner (grayed-out region) of Fig.~\ref{fig:sto}(b) is sufficiently
complicated that we chosen to provide a separate Fig.~\ref{fig:blowup}
to describe the behavior there.  The six panels of Fig.~\ref{fig:blowup}
show a blow-up of the phase diagram in the range
$D_z\in[0,0.05]$\,a.u., with color coding as explained in the caption.
We shall not
describe all the details here, as these delicate transitions
occur in quite a small region around the origin in $\bf D$ space
and are not very relevant to the broader discussion.

In Figs.~\ref{fig:sto}(a-d), the solid outer boundary (second
order transition to the rotationless phase) expands to larger
$D_x$ and $D_y$ with increasing $D_z$ (note the change of scale
from the first three to the last three panels).
In Fig.~\ref{fig:sto}(c), which is for $D_z$=0.08\,a.u.,
the first-order boundaries, shown by solid lines marked
by squares (magenta), are the remnant of the first-order
boundaries of Figs.~\ref{fig:blowup}(e-f); these diminish
and disappear as $D_z$ is increased further.  Then, by
the time $D_z$=0.16\,a.u.\ is reached in
Fig.~\ref{fig:sto}(d), a new pocket of rotationless phase
appears near the origin in the $D_x$-$D_y$ plane.  This
pocket grows until, at a critical value of
$D_z$=0.172\,a.u.\ shown in Fig.~\ref{fig:sto}(e),
the inner and outer regions connect and split the
region of rotational phases into four ellipses, as
shown for $D_z$=0.24\,a.u.\ shown in Fig.~\ref{fig:sto}(f).
We expected these ellipses to shrink and disappear with a
further increase of $D_z$, but in fact this happens only
very slowly; along the line $D_x=D_y=D_z$, the rotations
survive to quite large values of $\D$, as is confirmed by
the first-principles calculations upon which the model is
based.  We comment on this further is Sec.~\ref{sec:discuss}.

\subsubsection{Phase diagram for \PTO} \label{sec:pto-results}

\begin{figure*}[<+htpb+>]
  \begin{center}
    \includegraphics[height=1.93 in]{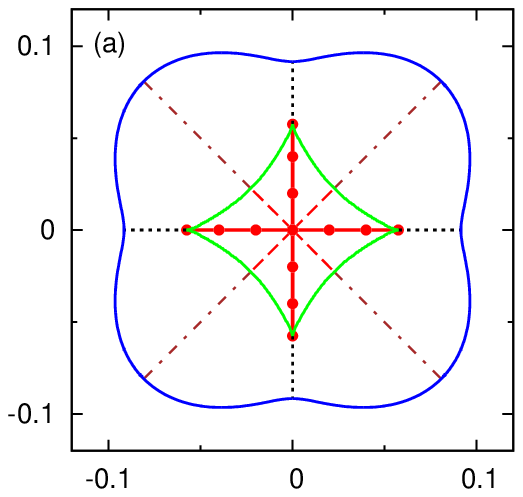}
    \includegraphics[height=1.93 in]{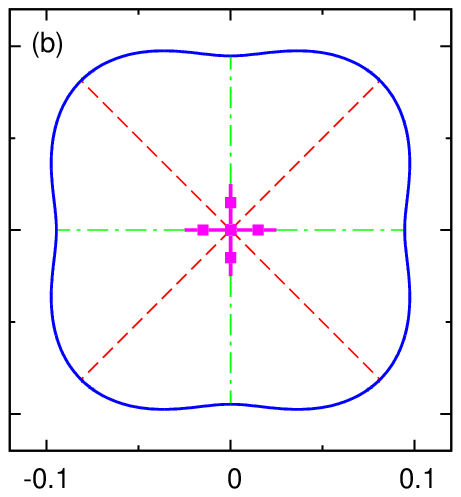}
    \includegraphics[height=1.93 in]{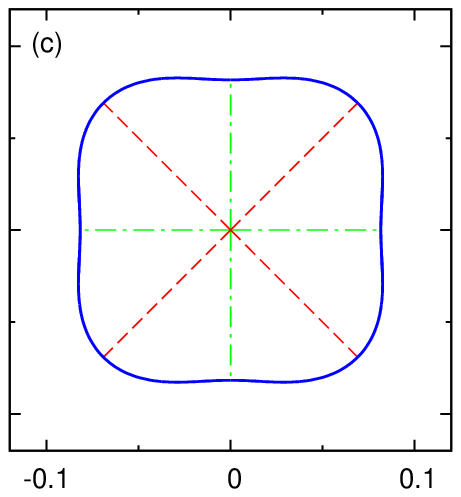}
  \end{center}
  \caption{Phase diagram for rotational phases of \PTO\ under
  applied $\D$ field.  $D_x$-$D_y$ are cuts plotted at
  (a) $D_z$=0.00 a.u.; (b) 0.02 a.u.; (c) 0.08 a.u.
  Conventions are similar to those of Fig.~\ref{fig:sto} (see
  text for details).}
  \label{fig:pto}
\end{figure*}

The situation is simpler for the rotational phase diagram of
\PTO.  The phase diagrams for several snapshots at increasing
$D_z$ are shown in Fig.~\ref{fig:pto} for \PTO\ using the same
conventions as in Fig.~\ref{fig:sto} wherever possible.
When $D_z$=0 as in Fig.~\ref{fig:pto}(a), the center point ($\D$=0)
is in the $\mathcal{R}$ phase.  The surrounding area enclosed by
the solid lines (green) is triclinic except along the [100]
and [110] symmetry lines, which are \Ma$_x$ (lines with circles,
red) and \Mb$_z$ (dashed lines, red)
respectively.  The transition is continuous across the
latter, but first-order across the former.  Essentially,
at small three-dimensional $\D$, the system prefers to be
in a slightly perturbed version of one of four $\mathcal{R}$
phases, depending on the octant in which $\D$ resides.
Using the notation ${\cal S}_{++-}$ denotes the octant with
$D_x>0$, $D_y>0$, $D_z<0$, etc., we find that the $\mathcal{R}$-like
phase with
$\hat{\theta}\simeq[111]$ is preferred in
${\cal S}_{+++}$ and
${\cal S}_{---}$;
$\hat{\theta}\simeq[\bar{1}11]$ is preferred in
${\cal S}_{-++}$ and
${\cal S}_{+--}$;
$\hat{\theta}\simeq[1\bar{1}1]$ is preferred in
${\cal S}_{+-+}$ and
${\cal S}_{-+-}$; and
$\hat{\theta}\simeq[11\bar{1}]$ is preferred in
${\cal S}_{++-}$ and
${\cal S}_{--+}$.
The planes $D_x$=0, $D_y$=0, and $D_z$=0 thus form
first-order boundaries in this small-$\D$ region, appearing as
solid lines (labeled with circles, red) in the 2D plots.
It follows that the octahedral
rotation can be ``switched'' between these $\mathcal{R}$-like
(actually, triclinic) phases by a small change of external electric
displacement field.

The area between the inner (green) and outer (blue) solid lines
in Fig.~\ref{fig:pto}(a)
is the $\mathcal{M}_{\rm Cz}$
phase, which becomes $\mathcal{O}_z$ along [110] directions
(short-dashed-dotted lines, brown) and $\mathcal{T}_x$ along [100]
direction (dotted lines, black).
The inner solid (green) lines
thus represent second-order phase boundaries at which
$\theta_z\rightarrow0$ as one passes to the outside.

As was the case for \STO, the high-symmetry phases for PTO\ in
Fig.~\ref{fig:pto}(a) become lower-symmetry phases as $D_z$
increases, Figs.~\ref{fig:pto}(b-c).  In fact, as soon as
$D_z>0$, the entire region inside the solid (blue) boundary
is generically triclinic.  Special cases occur along the dashed
lines (red), where the symmetry is \Ma\  or \Mb, and along
the dashed-dotted lines (green), which is \Mc.  When $D_z$
is small enough, as in Fig.~\ref{fig:pto}(b), the first-order
phase boundaries mentioned above are still visible as the
solid lines with squares (magenta) near the origin, corresponding to
the $\mathcal{O}_z$ phase, but with increasing $D_z$ these
shrink and then vanish, as shown in Fig.~\ref{fig:pto}(c). 
The solid (blue) boundary, outside which the rotational
phases disappear, can also be seen to shrink with increasing
$D_z$, at first slowly and then more rapidly, and to disappear
by the time $D_z$ reaches 0.10\,a.u.

\subsection{Discussion}
\label{sec:discuss}

There are quite significant differences between the rotational
phase diagrams for \STO\ and \PTO, as
shown in Figs.~\ref{fig:sto} and ~\ref{fig:pto}.  At small $\D$,
the major differences arise from the fact that the $\D$=0 ground
states are different, namely $\mathcal{T}$ and $\mathcal{R}$
respectively.  Thus, small applied $\D$ fields essentially
switch the system between $\mathcal{T}$-like phases in
\STO, or between $\mathcal{R}$-like phases in \PTO.

As $\D$ gets larger, the behavior becomes rather complex, but
we can identify an important difference that can be traced back
to the parameter values of the model.  Namely, we notice a much
more isotropic behavior of the outer boundary at which the rotations
disappear in \PTO\ compared to \STO.  In \PTO, for example, we
find that the critical magnitude of $D$ at which the rotations
disappear is $\sim$0.09\,a.u.\ and  $\sim$0.07\,a.u.\ in the
[100] and [111] $\D$-space directions respectively.
For \STO, on the other hand, the corresponding
values are $\sim$0.14 and $\sim$0.40\,a.u.\ respectively.
In addition to being larger (reflecting the stronger tendency to
rotational instability in \STO), the anisotropy between [100]
and [111] directions is very much greater, with rotations
extending much further in $\D$ space along the [111] direction.
This can be understood from the coefficients reported in Table I.
Restricting ourselves to the case of
$\D=(D_0,D_0,D_0)$ and $\bh=(\theta_0,\theta_0,\theta_0)$,
the critical displacement $D_{\rm c}$ can be obtained from
Eq.~\ref{eq:U} as $D_{\rm c}^2= -\mu / (\tau + 2 \lambda + \kappa)$.
 From this we obtain
$D_c$=0.34 and 0.07\,a.u.\ for \STO\ and \PTO\ respectively.
The greatly enhanced anisotropy and large value of $D_{\rm c}$ along
this [111] direction can thus be traced to the small value of
the denominator $(\tau + 2 \lambda + \kappa)$ for \STO.

It is useful to put the above results in perspective regarding
the dielectric behavior.  \STO\ is known experimentally to
remain paraelectric down to 0\,K, so that the entire space of
$\D$-fields should be accessible by varying an applied $\E$
field, even if the dielectric constant is very large.  \PTO,
on the other hand, is strongly ferroelectric, so that the
region of small $\D$ corresponds physically to the saddle
point of the multi-well energy landscape.  This internal-energy
landscaped as a function of $\D$ (without octahedral
rotations) was mapped out in our previous work~\cite{fixd},
where the
spontaneously polarized tetragonal ground state occurs at
$|D_{[001]}|=0.17$ a.u.  Similarly, the spontaneously polarized
states with constrained orthorhombic and rhombohedral symmetry
occur at $|D_{[110]}|$=0.15 a.u.\ and $|D_{[111]}|$=0.14 a.u.\
respectively. So, we can roughly think of this as a three-dimensional
``Mexican hat'' potential with a radius of $\sim$0.15\,a.u.  In
comparison, the results presented above show that
the octahedra rotations disappear for $|D_z|> 0.09$ a.u.\ and
$|D_{[110]}>0.11$ a.u.  Thus, the entire region of the
interesting rotational phase diagram shown in Fig.~\ref{fig:pto}
lies inside the Mexican-hat radius, in the region where the
crystal is unstable under fixed $\E$ (but not under fixed $\D$)
electric boundary conditions.

\section{Summary and conclusion}
\label{sec:summ}

In summary, we have investigated the phase transitions associated
with oxygen octahedral rotations in \STO\ and \PTO\ as a function
of a three-dimensional applied electric displacement field,
first directly from first-principles calculations and then also
using a fitted Landau-Devonshire model.
For \STO, the $\D$=0 ground state is tetragonal, with degenerate
states corresponding to the rotation angle lying along one of
the three Cartesian axes, and for small $\D$-vectors the ground
state is a weakly perturbed version of one of these states.
Similarly, for \PTO, the $\D$=0 ground state is rhombohedral,
with four degenerate states having rotation axis in one of the
[111] or related directions, and again a small $\D$ selects and
weakly perturbs one of these states.  However, as the strength
of $\D$ is increased, we find a quite complicated phase diagram
for each material, with both first- and second-order phase boundaries
appearing in different parts of the diagram.  The structure is
especially rich for the case of
\STO.  For both materials, the general
state associated with generic $D_x\ne D_y\ne D_z\ne 0$ is
triclinic, but states with higher symmetry tend to arise when
$\D$ itself has higher symmetry.  In both materials, the rotations
eventually disappear at sufficiently large values of applied $\D$.

Our work represents one of the first attempts to carry out a
systematic three-dimensional characterization of the interplay
between polar and octahedral-rotation degrees of freedom
in perovskites of this class.
While there is no external field that couples directly to
the rotational degrees of freedom, so that is very difficult
to find ways of controlling the rotations directly, the present work
suggests that such control should be possible indirectly via the
application of appropriate electric fields.
In any case, the observed richness of behavior suggests that
there may be much more to learn in other materials of this class
and in more distantly related materials.

\acknowledgments

This work was supported by ONR Grant N-00014-05-1-0054.
Computations were done at the Center for
Piezoelectrics by Design.


\begin{thebibliography}{99}

\bibitem{Nicole} Nicole A. Benedek and Craig J. Fennie, Phys.
  Rev. Lett. {\bf 106}, 107204 (2011).

\bibitem{spin} Jiawang Hong, Alessandro Stroppa, Jorge
  Iniguez, Silvia Picozzi, and David Vanderbilt, Phys. Rev. B
  {\bf 85}, 054417 (2012).

\bibitem{Zhong} W. Zhong and David Vanderbilt, Phys. Rev. Lett.
  {\bf 74}, 2587 (1995).

\bibitem{Sai} Na Sai and David Vanderbilt, Phys. Rev. B {\bf 62}, 13942 (2000).

\bibitem{Eklund} C.-J. Eklund, C. J. Fennie, and K. M. Rabe, Phys. Rev. B
  {\bf 79}, 220101 (2009).

\bibitem{Eric} Eric Bousquet, Matthew Dawber, Nicolas Stucki, Celine
  Lichtensteiger, Patrick Hermet, Stefano Gariglio, Jean-Marc
  Triscone, and Philippe Ghosez, Nature {\bf 452}, 732-736 (2008).

\bibitem{Ale} T. Fukushima, A. Stroppa, S. Picozzi, and J. M. Perez-Mato,
  Phys. Chem. Chem. Phys. {\bf 13}, 12186 (2011).

\bibitem{Jorge} Jorge Lopez-Prez and Jorge Iniguez, Phys. Rev. B
  {\bf 84}, 075121 (2011).

\bibitem{Claude} Claude Ederer and Nicola A. Spaldin, Phys. Rev. B 74,
  024102 (2006).

\bibitem{Lawes} Gavin Lawes, Physics {\bf 4}, 18 (2011).

\bibitem{Fennie}  Nicole A. Benedek, Andrew T. Mulder, and Craig
  J. Fennie,  J. Solid State Chem. {\bf 195}, 11 (2012).

\bibitem{James1} James M Rondinelli and Nicola A Spaldin, Adv.
  Mater.  {\bf 23}, 3363 (2011).

\bibitem{James2}James M. Rondinelli and Craig J. Fennie, Adv.
  Mater. {\bf 24}, 1961 (2012).

\bibitem{Ghosez1} Ph. Ghosez, E. Cockayne, U. V. Waghmare, and K. M. Rabe,
  Phys. Rev. B {\bf 60}, 836 (1999).

\bibitem{Ghosez2} P. Ghosez, D. Desquesnes, X. Gonze, K. M. Rabe,
  in R. E. Cohen (Ed.): Fundamental Physics of Ferroelectrics
  2000, AIP Conference Proceedings 535 (American Institute of
  Physics, Woodbury, New York 2000) pp.  102--110

\bibitem{Munkholm}  A. Munkholm, S. K. Streiffer, M. V. Ramana Murty,
  J. A. Eastman, Carol Thompson, O. Auciello, L. Thompson, J. F.
  Moore, and G. B. Stephenson, Phys. Rev. Lett. {\bf 88}, 016101
  (2001).

\bibitem{Karin}  Claudia Bungaro and K. M. Rabe, Phys. Rev. B {\bf
  71}, 035420 (2005).


\bibitem{James3} James M. Rondinelli and Sinisa Coh, Phys. Rev.
  Lett. {\bf 106}, 235502 (2011).

\bibitem{May} S. J. May, J.-W. Kim, J. M. Rondinelli, E. Karapetrova, N. A.
  Spaldin, A. Bhattacharya, and P. J. Ryan, Phys. Rev. B {\bf 82}, 014110
(2010).

\bibitem{Muller} K. A. Muller, W. Berlinger, and F. Waldner, Phys. Rev. Lett.
  {\bf 21}, 814 (1968).


\bibitem{Max-NM} M. Stengel, D. Vanderbilt and N.A. Spaldin,
Nat. Mater. {\bf 8}, 392 (2009)

\bibitem{fixd} Jiawang Hong and David Vanderbilt, Phys. Rev. B
  {\bf 84}, 115107 (2011).

\bibitem{david} David Vanderbilt and Morrel H. Cohen, Phys.
  Rev. B {\bf 63}, 094108 (2001).

\bibitem{note1}
Note that $\mathcal{T}$, $\mathcal{R}$ and $\mathcal{O}$ denote
only the symmetry of the octahedral rotation axis.  They do
not necessarily denote the symmetry of the overall phase,
which is determined also by the $\D$ field.  Real tetragonal,
rhombohedral and orthorhombic phases are denoted as T,
R and O, respectively, in this paper.


\bibitem{Perdew-Wang} J.P. Perdew and Y. Wang, Phys. Rev. B {\bf
45},13244 (1992)

\bibitem{Norm-conserve} N. Troullier and J.L. Martins, Phys. Rev. B
{\bf 43}, 1993 (1991)

\bibitem{MP} H.J. Monkhorst and J.D. Pack, Phys. Rev. B {\bf
13},5188 (1976)

\bibitem{Gonze} X. Gonze, {\it et al.}, Comp. Mat. Sci. {\bf 25}, 478 (2002).

\end{thebibliography}
\end{document}